\RequirePackage[2024-12-01]{latexrelease}
\documentclass[showpacs,showkeys,11pt,
preprint,preprintnumbers,nofootinbib,
groupedaddress,superscriptaddress,amsmath,amssymb]{revtex4}
\usepackage{xcolor}
\usepackage{tabularx}
\usepackage{amsfonts}
\usepackage{amsmath}
\usepackage{graphicx,subfigure}
\usepackage{caption}
\usepackage[export]{adjustbox}% http://ctan.org/pkg/adjustbox
\usepackage{verbatim} 
\usepackage{epsfig}
\usepackage{url}
\usepackage{booktabs}
\usepackage{multirow}
\usepackage{hhline}
\usepackage{feynmp}
\usepackage{csquotes}

\newcommand {\be}{\begin{equation}}
\newcommand {\ee}{\end{equation}}
\newcommand {\ba}{\begin{eqnarray}}
\newcommand {\ea}{\end{eqnarray}}

\begin{document}
\title{Probing the Scalar Sector: Discovery Reach for Heavy Higgs Pairs at a $\sqrt{s} = 6$~TeV Muon Collider in the 2HDM Alignment Limit}

%%%%%%%%%%%%%%%%%%%%%%%%%%%%%%%%%%%%%%%%%%%%%%%%%%%%%%%%%%%%%%%%%%%%%%%%%%%%%%%%
\author{Ijaz Ahmed}
\email{ijaz.ahmed@fuuast.edu.pk}
\affiliation{Federal Urdu University of Arts, Science and Technology, Islamabad, Pakistan}

\author{Muhammad Umar Farooq}
%\email{usmanahmed1661@gmail.com}
\affiliation{Riphah International, Islamabad, Pakistan}

\author{Farzana Ahmad}
\email{farzana@konkuk.ac.kr}
\affiliation{SERI, \& College of Engineering, Konkuk University, Seoul 05029, South Korea}

\author{Jamil Muhammad}
\email{mjamil@konkuk.ac.kr}
\affiliation{Sang-Ho College \& Department of Physics, Konkuk University, Seoul 05029, South Korea}
%%%%%%%%%%%%%%%%%%%%%%%%%%%%%%%%%%%%%%%%%%%%%%%%%%%%%%%%%%%%%%%%%%%%%%%%%%%%%%%%
\date{\today}

\begin{abstract}
This study provides a comprehensive phenomenological investigation into the discovery potential of heavy Higgs boson pairs ($HH, HA, AA, H^+H^-$) at a $\sqrt{s}=6$~TeV Muon Collider. Utilizing the Two-Higgs-Doublet Model (2HDM) Type-I within the alignment limit ($\sin(\beta-\alpha) \approx 1$), we evaluate two primary benchmarks with degenerate scalar masses of 1000~GeV (BP1) and 2000~GeV (BP2). Theoretical calculations performed reveal that Type-I branching fractions to third-generation fermions remain uniquely independent of $\tan\beta$, providing a stable signal across the investigated parameter space. We demonstrate that the Muon Collider environment allows for the precise identification of high-multiplicity hadronic final states. A key finding of this research is that the signal processes yield distinctive topological signatures: an 8-jet state ($4j+4b$) for charged pairs and a highly complex 12-jet state ($8j+4b$) for neutral pairs ($HA/AA$). These signatures, combined with hard transverse momentum distributions and central pseudorapidity ($|\eta| \le 3$), allow for nearly absolute suppression of Standard Model backgrounds like $t\bar{t}$, $W^+W^-Z$, and $ZZZ$. At an integrated luminosity of 10~ab$^{-1}$, we report a staggering statistical significance of 104,000 for the $H^+H^-$ channel and 3343 for the $HA$ channel in the BP1 scenario. Furthermore, total selection efficiencies were found to increase from approximately 20\% at BP1 to 47\% at BP2, suggesting that the decay products of heavier scalars are kinematically easier to resolve. We conclude that a 6~TeV Muon Collider offers an unparalleled discovery reach for the extended scalar sector, providing a definitive facility for probing physics beyond the Standard Model.
\end{abstract}

%\begin{document}
\maketitle
%%%%%%%%%%%%%%%%%%%%%%%%%%%%%%%%%%%%%%%%%%%%%%
\section{Introduction}
%%%%%%%%%%%%%%%%%%%%%%%%%%%%%%%%%%%%%%%%%%%%%%
The discovery of the Higgs boson at the Large Hadron Collider (LHC) in 2012 \cite{ATLAS:2012, CMS:2012} provided the final piece of the Standard Model (SM), confirming the mechanism of spontaneous electroweak symmetry breaking (EWSB) and the origin of fundamental particle masses \cite{Glashow:1961, Weinberg:1967}. Despite its success, the SM remains an incomplete description of the universe, failing to provide candidates for dark matter \cite{Peebles:2003}, account for the hierarchy problem \cite{Lykken:2010, Melo:2017}, or explain the observed matter-antimatter asymmetry. These deficiencies strongly suggest that the SM is an effective field theory valid up to an energy scale where new physics must emerge \cite{Gaillard:1999, Cowan:2012}.
Among the various proposed extensions, the Two-Higgs-Doublet Model (2HDM) stands as one of the most well-motivated and extensively studied extensions \cite{Branco:2012}. By introducing a second $SU(2)_L$ scalar doublet, the model predicts a rich scalar sector consisting of five physical states: two CP-even bosons ($h, H$), one CP-odd boson ($A$), and a pair of charged Higgs bosons ($H^\pm$) \cite{Botella:2018, Chakraborty:2015}. Current experimental data from the LHC has constrained the scalar sector close to the ``alignment limit,'' where the coupling profile of the light CP-even scalar $h$ becomes indistinguishable from that of the SM Higgs boson \cite{Enomoto:2016}. Consequently, searching for the heavier scalars ($H, A, H^\pm$) and understanding their self-couplings is paramount for probing the structure of the extended scalar potential \cite{Aoki:2009, Earman:2004}.

While the LHC and the proposed High-Luminosity LHC (HL-LHC) will continue to refine Higgs measurements, their reach is often limited by large QCD backgrounds and low signal-to-background ratios in certain BSM channels \cite{Baer:2013}. Future lepton colliders offer a much cleaner experimental environment. Recently, the multi-TeV Muon Collider has emerged as a frontrunner for the next generation of high-energy physics facilities \cite{Palmer:2014, Boscolo:2019, Accettura:2024, Capdevilla:2025}. Recently, the IMCC and US community have outlined a roadmap toward a 10 TeV facility \cite{Lucchesi:2025, FNAL:2024}. Since the muon is approximately 207 times more massive than the electron, synchrotron radiation is suppressed by a factor of $(m_\mu/m_e)^4 \approx 10^9$, allowing for circular acceleration to multi-TeV center-of-mass energies while maintaining high luminosity \cite{Franceschini:2021, Accettura:2023}. A 6 TeV muon collider effectively combines the precision of $e^+e^-$ machines with the high energy reach of hadron colliders, making it an ideal laboratory for probing heavy Higgs pair production \cite{Han:2021, IMCC:2024}.

The production of Higgs boson pairs ($HH, HA, AA, H^+H^-$) provides a direct probe into the scalar potential and the nature of the EWSB transition \cite{Sonmez:2018, Buttazzo:2018}. In the 2HDM framework, these cross-sections can be significantly enhanced compared to the SM, potentially within the reach of a high-energy muon collider \cite{DiLuzio:2019}. While this analysis focuses on the $s$-channel annihilation mechanism, which is highly effective at the $\sqrt{s} = 6$ TeV scale, it is important to note that the Vector Boson Fusion (VBF) contribution—as detailed by Han et al. (2021) \cite{Han:2021} and Buttazzo et al. (2018) \cite{Buttazzo:2018}—becomes increasingly significant at higher center-of-mass energies, eventually becoming the dominant production mode for heavy scalars in the multi-TeV regime \cite{Ma:2024, PhysRept:2024}. Accurate modeling of these processes requires sophisticated computational frameworks. Current analyses rely on automated tools such as \texttt{CalcHEP} \cite{Pukhov:2005} and \texttt{2HDMC} \cite{Eriksson:2010} for cross-section calculations, while event generation and kinematic analysis are typically performed via \texttt{MadGraph5\_aMC@NLO} \cite{Alwall:2014} and \texttt{MadAnalysis 5} \cite{Conte:2013}, often supplemented by symbolic manipulators like \texttt{FeynHelpers} \cite{Shtabovenko:2017}. 
In this work, we perform a detailed phenomenological study of heavy Higgs pair production at a multi-TeV Muon Collider operating at $\sqrt{s} = 6$ TeV. We evaluate the effective cross-sections and signal significance for the processes $\mu^+\mu^- \to HH, HA, AA, H^+H^-$ within the 2HDM alignment limit. Our analysis considers two primary benchmark points (BP1 and BP2) corresponding to degenerate Higgs masses of 1000 GeV and 2000 GeV, respectively. By simulating both signal and relevant SM backgrounds ($t\bar{t}$, $W^+W^-Z$, $ZZZ$) across integrated luminosities of 3.6 ab$^{-1}$ and 10 ab$^{-1}$, we demonstrate the discovery potential of this facility for the BSM scalar sector.
%%%%%%%%%%%%%%%%%%%%%%%%%%%%%%%%%%%%%%%%%%%%%%%%%%%%%%%%%%%%
\section{THEORETICAL FRAMEWORK}
%%%%%%%%%%%%%%%%%%%%%%%%%%%%%%%%%%%%%%%%%%%%%%%%%%%%%%%%%%
\subsection{2HDM Types using CalcHEP}
\begin{figure}[ht]
\centering
\includegraphics[width=0.65\textwidth]{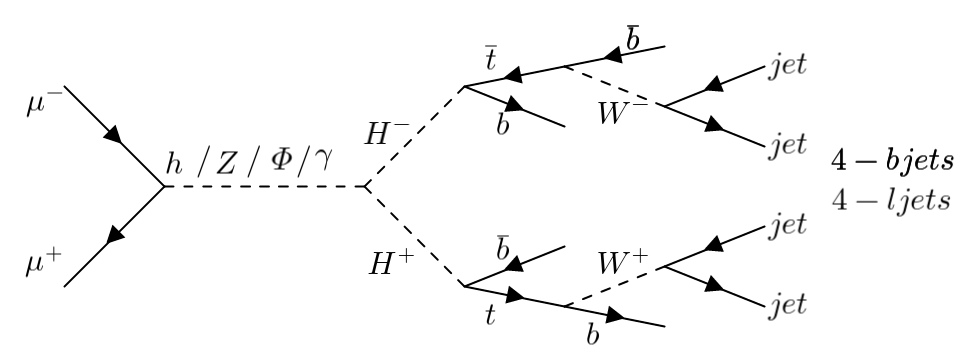}
 \caption{Branching ratio vs $\tan\beta$ for $H/A \to b\bar{b}, t\bar{t}, \tau^+\tau^-$ at $m_\phi = 2$ TeV.}
 \label{fig:5.5}
\end{figure}

\begin{figure}[ht]
\centering
 \includegraphics[width=0.65\textwidth]{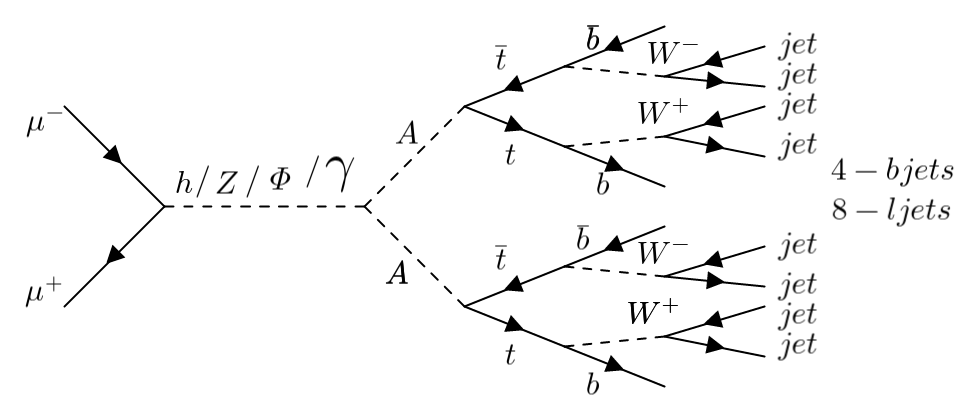}
\caption{Branching ratio vs $\tan\beta$ for $H^+ \to t\bar{b}$ and $H^+ \to \tau^+ \nu_\tau$ at $m_\phi = 2$ TeV.}
    \label{fig:5.6}
\end{figure}

%%%%%%%%%%%%%%%%%%%%%%%%%%%%%%%%%%%%%%%%%%%%%%%%%%%%%%%%%
In this work, CalcHEP is utilized to calculate and distinguish the branching fractions of neutral and charged Higgs decays into various fermion pairs across the four types of the Two Higgs Doublet Model (2HDM). While the pair production rates are identical across all 2HDM types due to their dependence on gauge couplings, the branching fractions provide unique signatures as they depend on the specific Yukawa couplings of each model. As illustrated in Figure \ref{fig:5.5}, the branching ratios for neutral Higgs decays ($H/A \to b\bar{b}, t\bar{t}, \tau^+\tau^-$) at a mass scale of 2 TeV demonstrate that Type-I remains entirely independent of $\tan\beta$. In contrast, for $\tan\beta > 5$, the other models show distinct dominance: the $H \to b\bar{b}$ decay mode is dominant in Type-II and IV, while the $H \to t\bar{t}$ and $H \to \tau^+\tau^-$ channels are significantly more prominent in Type-II and III compared to other configurations.
The distinction between 2HDM types is further clarified through the analysis of the charged Higgs sector. As depicted in Figure \ref{fig:5.6}, the branching fractions for $H^+ \to t\bar{b}$ and $H^+ \to \tau^+ \nu_\tau$ exhibit a strong sensitivity to the $\tan\beta$ parameter. While all 2HDM types exhibit similar behavior at low values ($\tan\beta < 5$), they diverge significantly in the high $\tan\beta$ region, allowing for the separation of Type-II and III from the others. Specifically, the $H^+ \to t\bar{b}$ channel is found to be the dominant decay mode in Type-I and IV, whereas the $H^+ \to \tau^+ \nu_\tau$ mode becomes the presiding decay mechanism in Type-II and III. These variations highlight the potential for a multi-TeV muon collider to determine the specific type of the 2HDM through the measurement of these distinct fermionic decay signatures.
%%%%%%%%%%%%%%%%%%%%%%%%%%%%%%%%%%%%%%%%%%%%%%%
% Figures placeholders
%%%%%%%%%%%%%%%%%%%%%%%%%%%%%%%%%%%%%%%%%%%%%%%%%%%%%5
%%%%%%%%%%%%%%%%%%%%%%%%%%%%%%%%%%%%%%%%%%%%%%%%%%%%%%%%%
%%%%%%%%%%%%%%%%%%%%%%%%%%%%%%%%%%%%%%%%%%%%%%%%%%%%%%%%%%%%%%
%%%%%%%%%%%%%%%%%%%%%%%%%%%%%%%%%%%%%%%%%%%%%%%%%%%%%%%%%%%%%%5
\subsection{The Two-Higgs-Doublet Model (2HDM)}
%%%%%%%%%%%%%%%%%%%%%%%%%%%%%%%%%%%%%%%%%%%%%%%%%%%%%%%%%%%%%%5
The 2HDM extends the SM by introducing a second complex scalar doublet. Both doublets $\Phi_1$ and $\Phi_2$ possess the same hypercharge $Y = 1$ \cite{Branco:2012}. This extension results in five physical scalar states after EWSB:
\begin{itemize}
    \item Two CP-even neutral scalars ($h, H$), where $h$ is typically identified as the 125 GeV SM-like Higgs.
    \item One CP-odd neutral scalar ($A$).
    \item Two charged Higgs bosons ($H^\pm$).
\end{itemize}

The scalar potential for a CP-conserving 2HDM with a softly broken $Z_2$ symmetry is expressed as \cite{Branco:2012}:
\begin{equation}
\begin{split}
V(\Phi_1, \Phi_2) = & m_{11}^2\Phi_1^\dagger\Phi_1 + m_{22}^2\Phi_2^\dagger\Phi_2 - [m_{12}^2\Phi_1^\dagger\Phi_2 + \text{h.c.}] \\
& + \frac{1}{2}\lambda_1(\Phi_1^\dagger\Phi_1)^2 + \frac{1}{2}\lambda_2(\Phi_2^\dagger\Phi_2)^2 + \lambda_3(\Phi_1^\dagger\Phi_1)(\Phi_2^\dagger\Phi_2) \\
& + \lambda_4(\Phi_1^\dagger\Phi_2)(\Phi_2^\dagger\Phi_1) + \frac{1}{2}\lambda_5[(\Phi_1^\dagger\Phi_2)^2 + \text{h.c.}]
\end{split}
\end{equation}

Two key parameters define the phenomenology of this model: the ratio of the two VEVs, $\tan\beta = v_2/v_1$, and the mixing angle $\alpha$ between the neutral CP-even states \cite{Dubinin:2025}. In the ``alignment limit,'' defined by $\sin(\beta - \alpha) \to 1$, the properties of the light scalar $h$ coincide with the SM Higgs boson \cite{Enomoto:2016}. Current global fits have significantly constrained the A2HDM parameter space \cite{Karan:2024}. To avoid Flavour Changing Neutral Currents (FCNC) at the tree level, discrete $Z_2$ symmetries are imposed, leading to four distinct types of 2HDM summarized in Table II \cite{Aoki:2009, Hashemi:2024}.
%%%%%%%%%%%%%%%%%%%%%%%%%%%%%%%%%%%%%%%%%%%%%%%%%%%%%%%%%%%%%%
\section{SIMULATION METHODOLOGY AND MUON COLLIDER SETUP}
%%%%%%%%%%%%%%%%%%%%%%%%%%%%%%%%%%%%%%%%%%%%%%%%%%%%%%%%%%%%%%
\subsection{The Multi-TeV Muon Collider Environment}
%%%%%%%%%%%%%%%%%%%%%%%%%%%%%%%%%%%%%%%%%%%%%%%%%%%%%%%%%%%%%%5
The Muon Collider represents a unique high-energy frontier facility that combines the advantages of both hadron and electron machines. Unlike protons, muons are fundamental particles, ensuring that the full center-of-mass energy ($\sqrt{s}$) is available for the hard scattering process \cite{Palmer:2014}. Furthermore, the power loss through synchrotron radiation for a particle of mass $m$ and energy $E$ scales as $1/m^4$. For a muon collider, this suppression relative to an electron machine is given by:
\begin{equation}
\frac{P_{\text{loss}}(\mu)}{P_{\text{loss}}(e)} = \left( \frac{m_e}{m_\mu} \right)^4 \approx 10^{-9}
\end{equation}
This massive reduction in energy loss allows for circular acceleration to multi-TeV scales within a compact footprint \cite{Franceschini:2021}. Recent design challenges regarding siting at major labs like Fermilab have been addressed in the latest roadmap \cite{Metral:2025, Eldred:2025}.
%%%%%%%%%%%%%%%%%%%%%%%%%%%%%%%%%%%%%%%%%%%%%%%%%%%%%%%%%
\subsection{Computational Framework and Event Generation}
%%%%%%%%%%%%%%%%%%%%%%%%%%%%%%%%%%%%%%%%%%%%%%%%%%%%%%%%%
Numerical analysis is performed using a multi-stage simulation pipeline. Theoretical consistency—including vacuum stability and unitarity—is verified using \texttt{2HDMC 1.8.0} \cite{Eriksson:2010}. Signal and background events are generated using \texttt{MadGraph5\_aMC@NLO} \cite{Alwall:2014}. The production cross-sections for signal processes:
\begin{equation}
\mu^+\mu^- \to HH, \quad \mu^+\mu^- \to HA, \quad \mu^+\mu^- \to AA, \quad \mu^+\mu^- \to H^+H^-
\end{equation}
are calculated at the parton level, including the analysis of multi-Higgs production via photon and vector boson fusion \cite{EPJC:2024, PhysRept:2024}. For the background analysis, we consider dominant SM processes with similar final-states, specifically $t\bar{t}$, $W^+W^-Z$, and $ZZZ$ \cite{Buttazzo:2018, Zuliani:2024}.

\begin{figure}[ht]
    \centering
    % If you are combining two images into one figure
     \includegraphics[width=8cm,height=5.5cm]{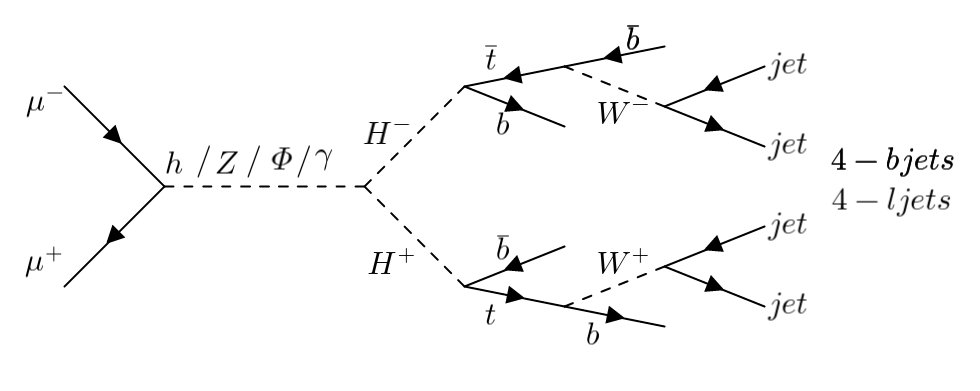} 
     \includegraphics[width=8cm,height=5.5cm]{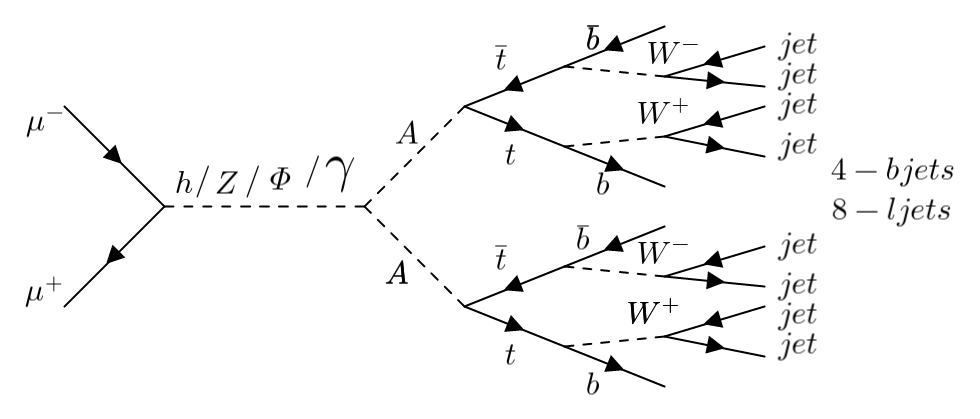}
    \caption{Representative Feynman diagrams for the signal processes $\mu^+\mu^- \to H^+H^-$ (left) and $\mu^+\mu^- \to AA$ (right) at a multi-TeV Muon Collider. These diagrams illustrate the complex decay chains through top quarks and gauge bosons, which result in the high-multiplicity hadronic final states ($N_{jets} \ge 8$ for neutral pairs) utilized for effective background suppression.}
    \label{fig:feynman_signals}
\end{figure}

The leading-order (LO) Feynman diagrams for the primary signal processes, 
$\mu^+\mu^- \to H^+H^-$ and $\mu^+\mu^- \to HA$, are depicted in 
Fig.~\ref{fig:feynman_signals}. These diagrams illustrate the 
$s$-channel annihilation mechanism and the subsequent decay chains 
leading to top-quark pairs, which were modeled using \texttt{MadGraph5\_aMC@NLO}.
%%%%%%%%%%%%%%%%%%%%%%%%%%%%%%%%%%%%%%%%%%%%%%%%%%%%%%%%
%%%%%%%%%%%%%%%%%%%%%%%%%%%%%%%%%%%%%%%%%%%%%%%%%%%%%%%%
Heavy Higgs pair production through $s$-channel annihilation yields distinctive signal topologies 
characterized by high jet multiplicities. The leading-order (LO) decay chains for the 
charged Higgs pair process ($\mu^+\mu^- \to H^+H^-$) and the CP-odd neutral pair 
process ($\mu^+\mu^- \to AA$) are illustrated in Figs.~4 and 5, respectively. 
Specifically, the $H^+H^-$ channel produces an eight-jet final state ($4j + 4b$) 
via the hadronic decays of top quarks and $W^\pm$ bosons. In contrast, the neutral 
$AA$ channel---and similarly the $HH$ process---results in a more complex twelve-jet 
signature ($8j + 4b$), as shown in Fig.~5. These high-multiplicity hadronic final 
states are a defining hallmark of signal processes within the 2HDM framework at 
multi-TeV scales. Such signatures provide a robust topological discriminator, 
enabling near-complete suppression of Standard Model (SM) backgrounds like $t\bar{t}$ 
and $VVZ$, which are typically characterized by lower jet counts and distinct 
kinematic profiles.

% --- Figure Code ---
\begin{figure}[ht]
    \centering
    \includegraphics[width=0.65\textwidth]{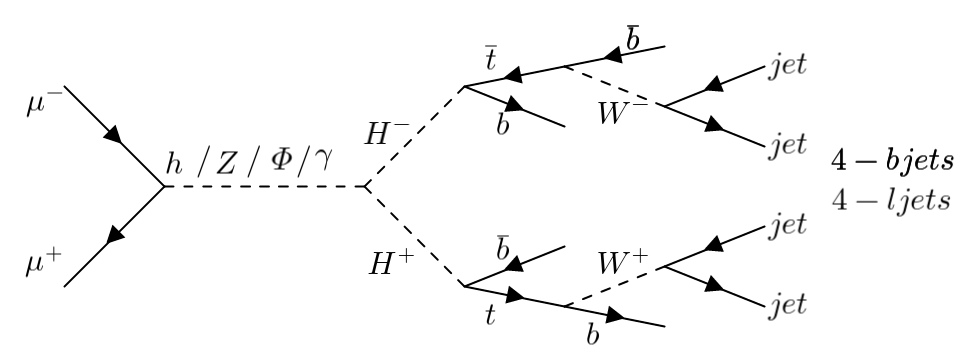} 
    \caption{Full leading-order Feynman diagram for the signal process $\mu^+\mu^- \to H^+H^-$, illustrating the decay chain into an 8-jet final state ($4j+4b$).}
    \label{fig:HpHm_full}
\end{figure}
%%%%%%%%%%%%%%%%%%%%%%%%%%%%%%%%%%%%%%%%%%
\begin{figure}[ht]
    \centering
    \includegraphics[width=0.65\textwidth]{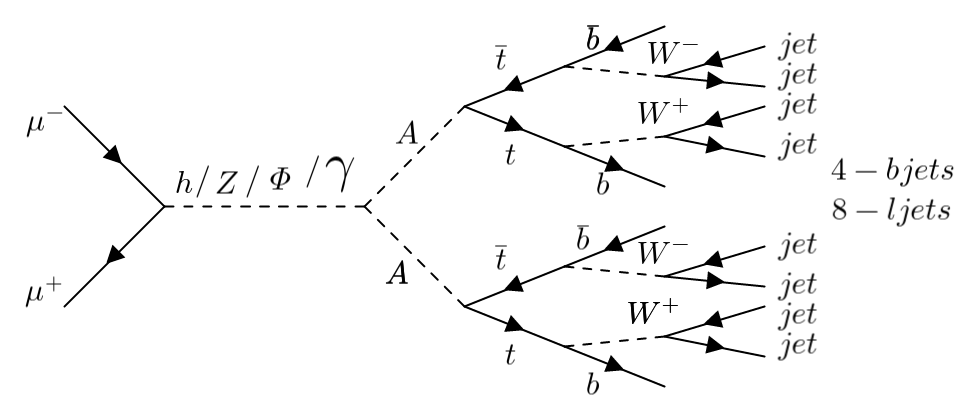} 
    \caption{Feynman diagram for the CP-odd neutral Higgs pair production $\mu^+\mu^- \to AA$, demonstrating the high-multiplicity 12-jet final state ($8j+4b$) resulting from four top quarks.}
    \label{fig:AA_full}
\end{figure}
%%%%%%%%%%%%%%%%%%%%%%%%%%%%%%%%%%%%%%%%%%%%%%
\subsection{Event Selection and Kinematic Cuts}
%%%%%%%%%%%%%%%%%%%%%%%%%%%%%%%%%%%%%%%%%%%%%%
To optimize the signal-to-background ratio ($S/B$), we implement a series of kinematic selection cuts using \texttt{MadAnalysis 5} 
To maximize the signal significance and optimize the signal-to-background ratio ($S/B$), we apply a sequence of kinematic selection cuts tailored to the high-multiplicity final states characteristic of heavy scalar decays \cite{Conte:2013}. These cuts are designed to exploit the distinct topology of heavy scalar decays, which typically result in high-multiplicity jet final states. The primary variables utilized include:
\begin{itemize}
    \item \textbf{Transverse Momentum ($P_T$):} Ensuring high momentum transfer for the reconstructed jets.
    \item \textbf{Pseudo-rapidity ($\eta$):} Restricting jets to the central region of the detector ($|\eta| \leq 3$) to reduce beam-induced backgrounds.
    \item \textbf{Separation ($\Delta R$):} Utilizing a cone size of $\Delta R = 0.2$ for jet clustering and separation \cite{Shtabovenko:2017}.
\end{itemize}

The specific selection criteria applied to all generated events are summarized in Table \ref{tab:cuts}.

\begin{table}[h]
\centering
\caption{Kinematic selection cuts applied for signal and background discrimination.}
\label{tab:cuts}
\begin{tabular}{|l|c|}
\hline
\toprule
\textbf{Variable} & \textbf{Selection Criteria} \\  \hline
\midrule
Transverse Momentum ($P_T^{jet}$) & $\geq 10$ GeV \\ 
Pseudo-rapidity ($\eta^{jet}$) & $|\eta| \leq 3$ \\ 
Jet Separation ($\Delta R$) & $0.2$ \\ 
Jet Multiplicity ($N_{jets}$) & $\geq 8$ (Neutral) / $\geq 4$ (Charged) \\ \hline
\bottomrule
\end{tabular}
\end{table}
%%%%%%%%%%%%%%%%%%%%%%%%%%%%%%%%%%%%%%%%%%%%%%%%%%%%%%%%
\section{Results and Discussion}
%%%%%%%%%%%%%%%%%%%%%%%%%%%%%%%%%%%%%%%%%%%%%%%%%%%%%%%%%%
\subsection{Production Cross-sections and Benchmark Analysis}
%%%%%%%%%%%%%%%%%%%%%%%%%%%%%%%%%%%%%%%%%%%%%%%%%%%%%%%%%%
The investigation begins with the evaluation of the production cross-sections for heavy Higgs pairs at the $\sqrt{s} = 6$ TeV Muon Collider. Following the theoretical constraints of the 2HDM Type-I in the alignment limit, we analyze two primary benchmark points (BP1 and BP2), as defined in Table II. 
%%%%%%%%%%%%%%%%%%%%%%%%%%%%%%%%%%%%%%%%%%%%%%%%%%%%%%%%%%
\begin{table}[h!]
\centering
\caption{Benchmark points for the 2HDM scalar sector at a 6~TeV Muon Collider \cite{Eriksson:2010, Han:2021}.}
\label{tab:benchmarks_6_1}
\begin{tabular}{|l|c|c|c|c|c|c|}
\hline
\toprule
\textbf{Point} & $m_h$ (GeV) & $m_H$ (GeV) & $m_A$ (GeV) & $m_{H^\pm}$ (GeV) & $\sqrt{s}$ (TeV) & $L_{int}$ ($fb^{-1}$) \\ 
\hline
\midrule
BP1 & 125 & 1000 & 1000 & 1000 & 6 & 3.6/10 ($ab^{-1}$)  \\ 
BP2 & 125 & 2000 & 2000 & 2000 & 6 & 3.6/10 ($ab^{-1}$) \\ \hline
\bottomrule
\end{tabular}
\end{table}
%%%%%%%%%%%%%%%%%%%%%%%%%%%%%%%%%%%%%%%%%%%%%%%%%%%%%%%%%%
%Fig. 1~\ref{fig:cross_vs_sqrts}
%Fig.2 ~\ref{fig:cross_vs_mass}) 
The behavior of the cross-section ($\sigma$) relative to the center-of-mass energy and scalar mass (illustrated in Fig. 6 and Fig. 7) reveals critical threshold effects. Below the $2m_\Phi$ threshold, the cross-sections are vanishingly small due to kinematic phase-space suppression. Upon reaching the threshold, the $s$-channel resonant structure induces a sharp increase in $\sigma$. Subsequently, the cross-section follows a predictable $1/s$ scaling behavior characteristic of high-energy annihilation, resulting in a gradual decline at higher energies \cite{Han:2021}.

For the BP2 scenario, the increased mass leads to a significantly lower peak cross-section compared to BP1. This is because the production of heavier scalars requires higher energy density and is further constrained by a reduced phase-space volume~\cite{Buttazzo:2018}. These results underscore that while the Muon Collider is a precision machine, its discovery reach is fundamentally governed by the proximity of the center-of-mass energy to the production threshold of the new scalar states.
The expected event yields for the signal processes at the second benchmark point (BP2) are detailed in Table \ref{tab:events_BP2}. The data highlights that the charged Higgs pair production ($H^+H^-$) is significantly more abundant than neutral pairs at higher masses. The high branching ratios to top-quark pairs across all channels facilitate a robust signature for detection.
%%%%%%%%%%%%%%%%%%%%%%%%%%%%%%%%%%%%%%%%%%%%%%%%%%%%%%%%%%%%%%%%%%%%
\begin{table}[ht]
\centering
\caption{Number of events for all signal processes at BP2 at $L_{int} = 3600 \, \text{fb}^{-1}$.}
\begin{tabular}{|l|c|c|c|c|c|c|}
\hline
Signal process & $\sigma$(pb) & Br($H \to t\bar{t}$) & Br($A \to t\bar{t}$) & $W^{\pm}/Z \to jj$ & No. of events \\ \hline
$\mu^+\mu^- \to H^+H^-$ & 6.511 & - & - & $(0.7)^2$ & $1.1 \times 10^7$ \\
$\mu^+\mu^- \to HH$ & 0.0039 & $(0.9988)^2$ & - & $(0.7)^4$ & $3.4 \times 10^3$ \\
$\mu^+\mu^- \to AA$ & 0.0323 & - & $(0.9986)^2$ & $(0.7)^4$ & $2.8 \times 10^4$ \\
$\mu^+\mu^- \to HA$ & 0.3247 & 0.9988 & 0.9986 & $(0.7)^4$ & $2.8 \times 10^5$ \\ \hline
\end{tabular}
\label{tab:events_BP2}
\end{table}
%%%%%%%%%%%%%%%%%%%%%%%%%%%%%%%%%%%%%%%%%%%%%%%%%%%%%%%%%%%%%%%%%%%%
% --- Composite Figure: Cross Section Analysis ---
\begin{figure}[htbp]
    \centering
    \begin{minipage}{0.45\textwidth}
        \centering
        \includegraphics[width=\textwidth]{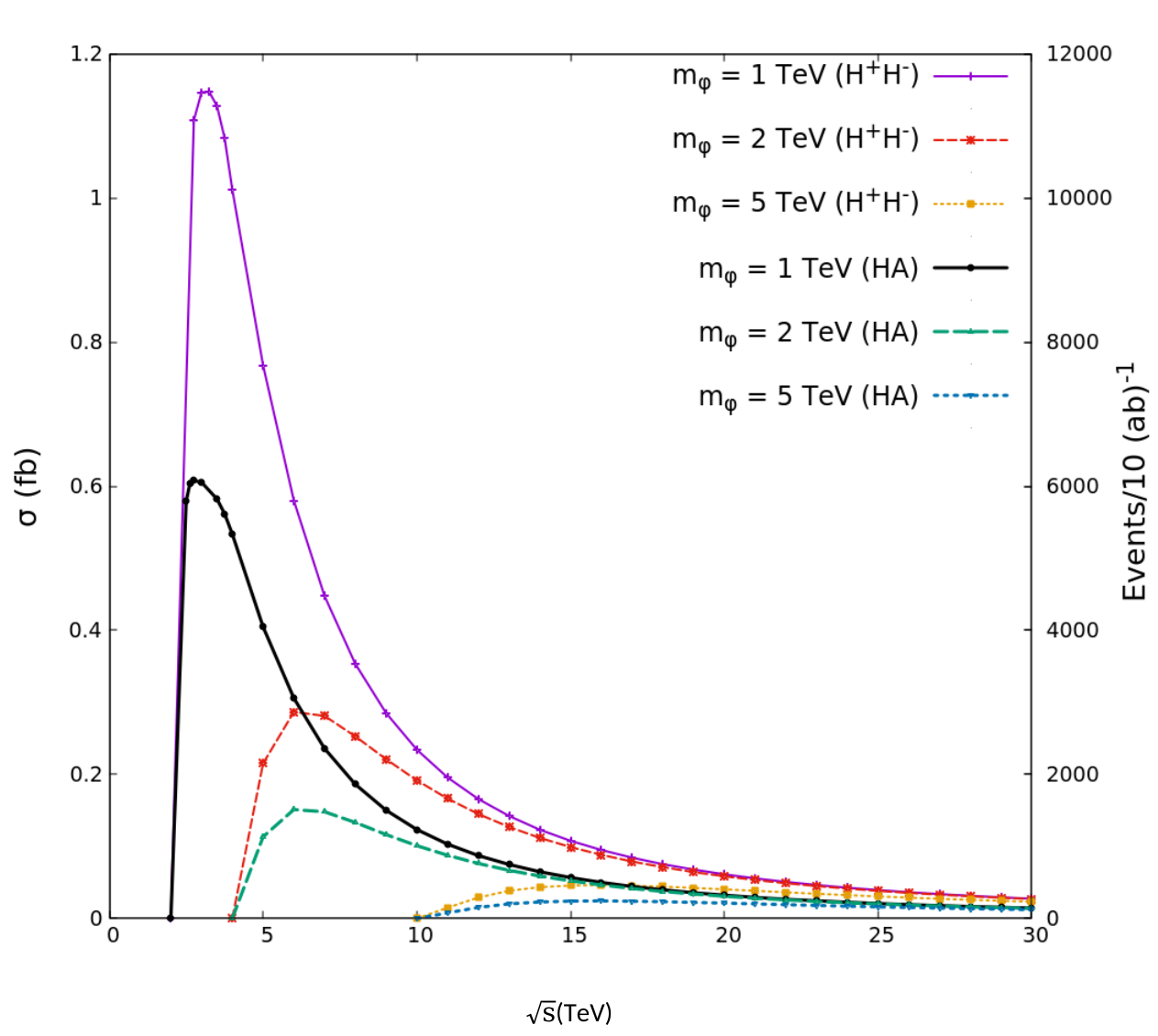} % Replace with actual filename
        \caption{Cross-section versus $\sqrt{s}$ at various degenerate Higgs masses.}
        \label{fig:cross_energy}
    \end{minipage}
    \hfill
    \begin{minipage}{0.48\textwidth}
        \centering
        \includegraphics[width=\textwidth]{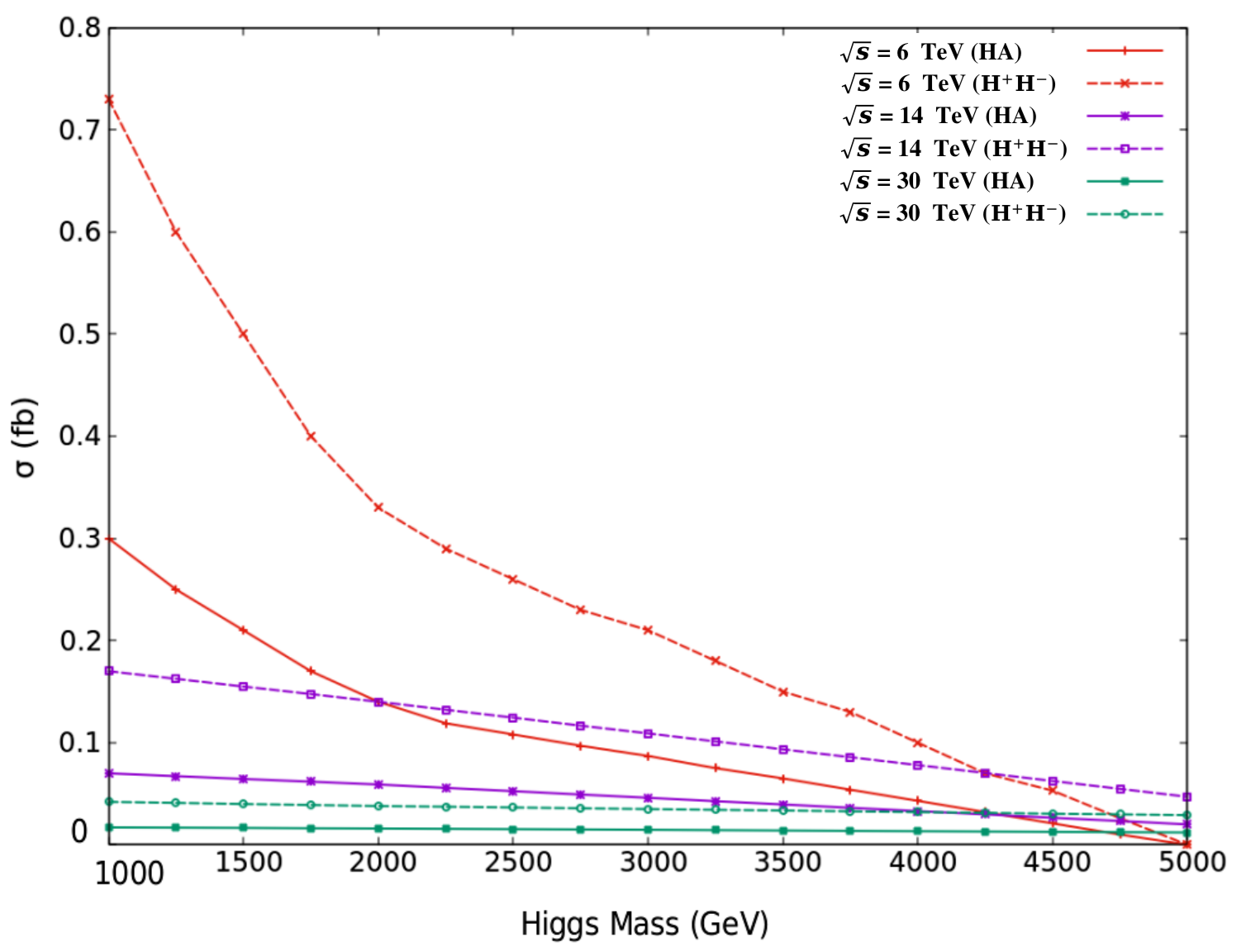} % Replace with actual filename
        \caption{Cross-section versus degenerate Higgs mass at different center of mass energies.}
        \label{fig:cross_mass}
    \end{minipage}
\end{figure}
%%%%%%%%%%%%%%%%%%%%%%%%%%%%%%%%%%%%%%%%%%%%%%%%%%%%%%%%%%%%%%%%%%%%
The cross-sections for the dominant Standard Model background processes at the center-of-mass energy of 6 TeV are summarized in Table \ref{tab:backgrounds}.
 The Standard Model backgrounds, specifically the $ZZZ$ channel, are found to be several orders of magnitude smaller than the signal cross-sections. This inherent suppression provides an advantageous signal-to-background ratio prior to kinematic refinement.
%%%%%%%%%%%%%%%%%%%%%%%%%%%%%%%%%%%%%%%%%%
\begin{table}[ht]
\centering
\caption{Standard Model Backgrounds cross section at $\sqrt{s} = 6$ TeV.}
\begin{tabular}{|l|c|c|c|}
\hline
Process & $W^+W^-Z$ & $t\bar{t}$ & $ZZZ$ \\ \hline
Cross Section (pb) & $3.859 \times 10^{-3}$ & $1.953 \times 10^{-3}$ & $2.429 \times 10^{-5}$ \\ \hline
\end{tabular}
\label{tab:backgrounds}
\end{table}
%%%%%%%%%%%%%%%%%%%%%%%%%%%%%%%%%%%%%%%%%%%%%%%%%%%%%%%%%%
\subsection{Kinematic Distributions and Background Rejection}
%%%%%%%%%%%%%%%%%%%%%%%%%%%%%%%%%%%%%%%%%%%%%%%%%%%%%%%%%%
To distinguish the signals ($\mu^+\mu^- \to HH, HA, AA, H^+H^-$) from the dominant SM backgrounds ($t\bar{t}, W^+W^-Z, ZZZ$), we analyzed the distributions of transverse momentum ($P_T$), pseudo-rapidity ($\eta$), and angular separation ($\Delta R$) in Figures 8 and 9.

\begin{figure}[ht]
    \centering
    \includegraphics[width=1.0\textwidth]{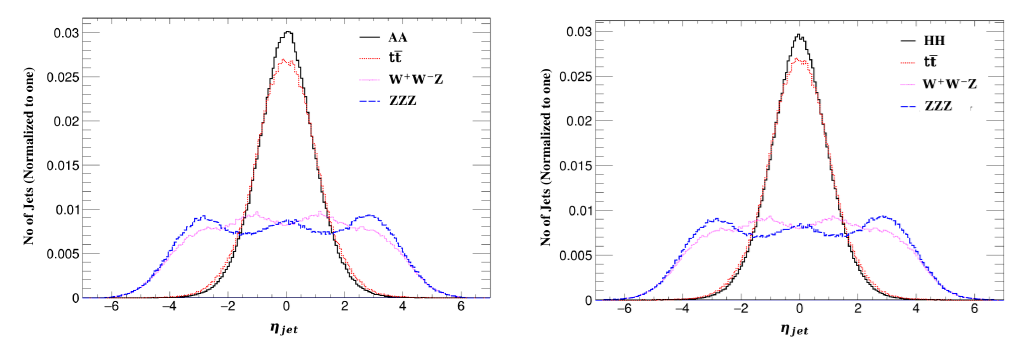} 
       \caption{Jets Pseudorapidity ($\eta$) distributions for the neutral Higgs signals $\mu^+\mu^- \to AA$ and $\mu^+\mu^- \to HH$ compared with Standard Model background processes. }
    \label{fig:eta_neutral}
\end{figure}

\begin{figure}[ht]
    \centering
\includegraphics[width=1.0\textwidth]{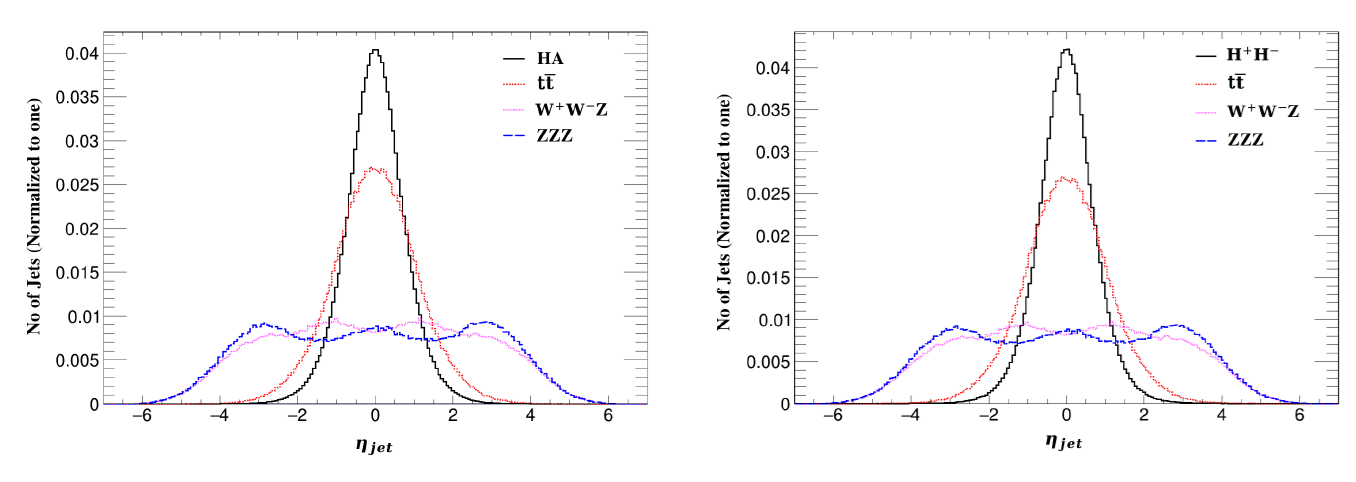} 
    \caption{Jets Pseudorapidity ($\eta$) distributions for signal processes $\mu^+\mu^- \to HA$ and $\mu^+\mu^- \to H^+H^-$ compared with Standard Model background processes.}
    \label{fig:eta_charged}
\end{figure}

The $P_T$ distributions (Figure 10) show that signal jets are significantly harder than background jets. This momentum shift is attributed to the heavy scalars decaying into high-energy top quarks and vector bosons. Consequently, the application of a $P_T \geq 10$~GeV cut serves as a powerful discriminator \cite{Shtabovenko:2017}. Similarly, the $\eta$ distributions (Figure 8 and 9) demonstrate signal centrality ($|\eta| \leq 3$), which contrasts with the forward-peaked nature of many beam-induced backgrounds. The angular separation $\Delta R \approx 0.2$ observed in Figures 12 indicates that while the final state is crowded, the jets are sufficiently resolved for accurate $b$-tagging \cite{Pukhov:2005}.
To further differentiate the 2HDM signal from the Standard Model (SM) noise, the angular distribution of the reconstructed jets is analyzed. The pseudorapidity ($\eta$) distributions for the neutral Higgs pairs ($AA$ and $HH$) and the mixed/charged pairs ($HA$ and $H^+H^-$) are shown in Fig. \ref{fig:eta_neutral} and Fig. \ref{fig:eta_charged}, respectively. The results demonstrate that jets originating from the decay of heavy Higgs bosons are highly central, with a sharp peak centered at $\eta \approx 0$. This behavior is expected for the pair production of massive scalars at a 6 TeV collider, where the particles are produced with a relatively low longitudinal boost.
Conversely, the SM background processes—specifically $t\bar{t}$ and electroweak diboson channels—exhibit much broader and flatter distributions, with a significant number of jets extending into the forward regions of the detector. This distinct contrast in angular topology confirms that the kinematic requirement of $|\eta| \leq 3$ is highly effective at preserving the signal while suppressing a large fraction of the SM background and beam-induced interference.
%%%%%%%%%%%%%%%%%%%%%%%%%%%%%%%%%%%%%%%%%%%%%%%%%%%%%%%%%%%
%%%%%%%%%%%%%%%%%%%%%%%%%%%%%%%%%%%%%%%%%%%%%%%%%%%%%%%%%%%
%%%%%%%%%%%%%%%%%%%%%%%%%%%%%%%%%%%%%%%%%%%%%%%%%%%%%%%%%%%
%%%%%%%%%%%%%%%%%%%%%%%%%%%%%%%%%%%%%%%%%%%%%%%%%%%%%%%%%%%%%%%%%%%%
% --- Composite Figure: Kinematics (PT and Multiplicity) ---
\begin{figure}[htbp]
    \centering
    \includegraphics[width=1.0\textwidth]{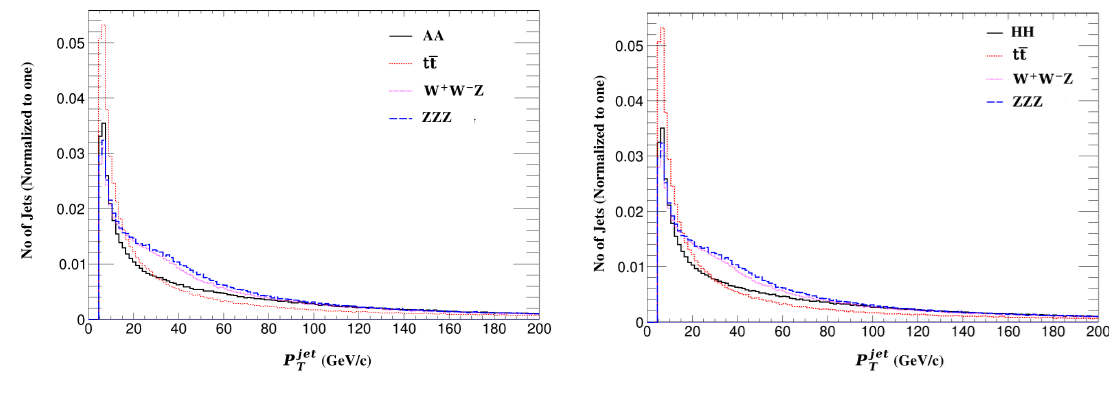} % AA/HH PT
    \includegraphics[width=1.0\textwidth]{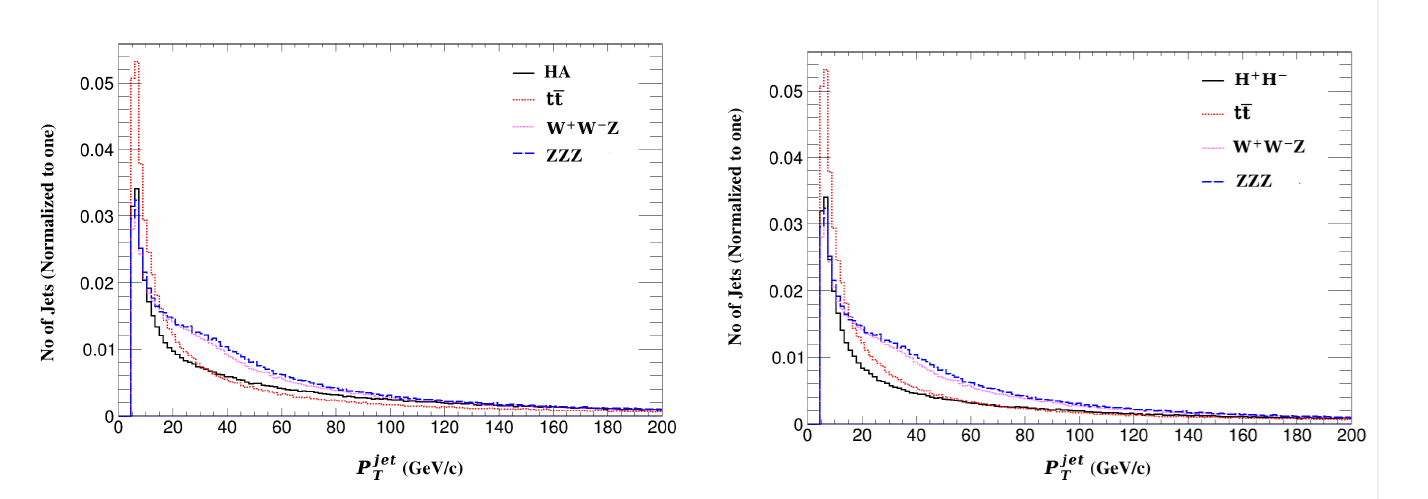} % HA/H+H- PT
    \caption{Transverse momentum distributions of jets for neutral (left) and charged (right) Higgs signal processes compared to Standard Model backgrounds.}
    \label{fig:pt_dist}
\end{figure}
%%%%%%%%%%%%%%%%%%%%%%%%%%%%%%%%%%%%%%%%%%%%%%%%%%%%%%%%%%%%%%%%%%%%
\begin{figure}[htbp]
    \centering
    \includegraphics[width=1.0\textwidth]{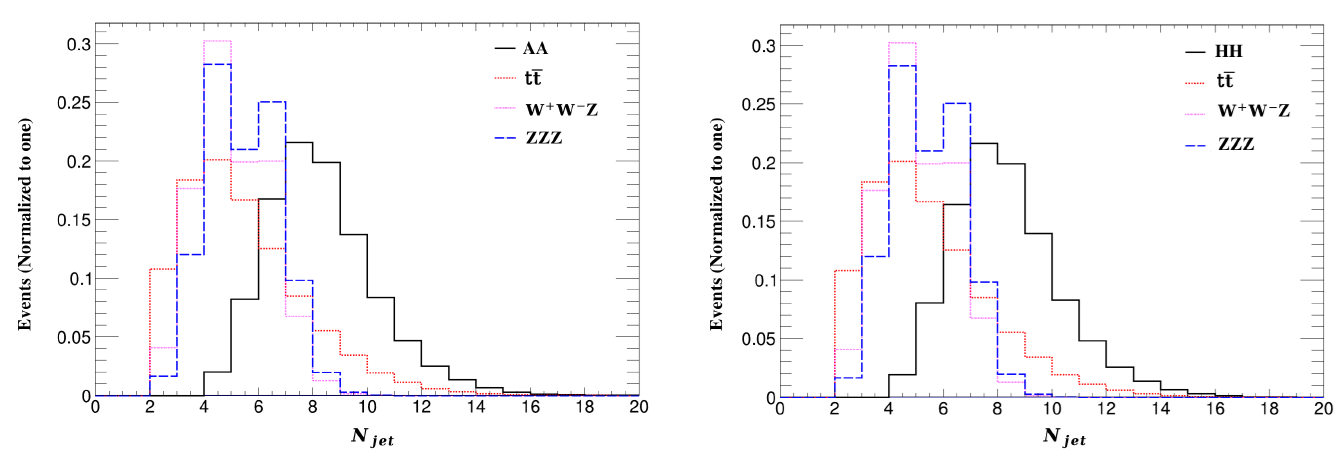} % N-jet Neutral
    \includegraphics[width=1.0\textwidth]{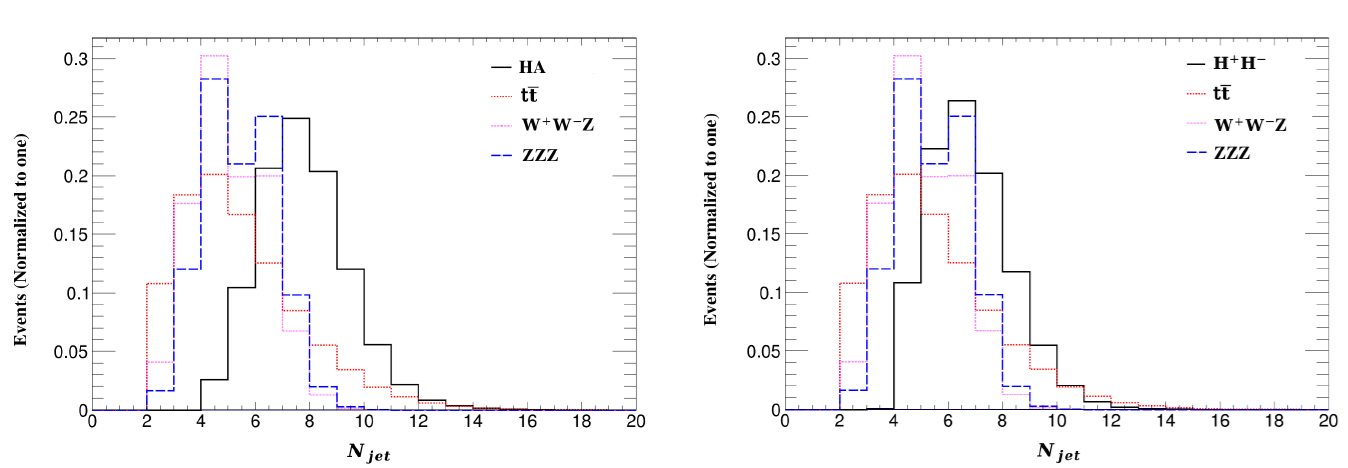} % N-jet Charged
    \caption{Distribution of light jet multiplicity ($N_{jets}$) for signals and SM backgrounds, demonstrating the high-multiplicity threshold used for signal selection.}
    \label{fig:njet_dist}
\end{figure}
%%%%%%%%%%%%%%%%%%%%%%%%%%%%%%%%%%%%%%%%%%%%%%%%%%%%%%%%%%%%%%%%%%%%
\subsection{Selection Efficiencies and Multiplicity}
%%%%%%%%%%%%%%%%%%%%%%%%%%%%%%%%%%%%%%%%%%%%%%%%%%%%%%%%%%
The complex decay chains of heavy Higgs bosons in the 2HDM lead to high jet multiplicities ($N_{jets}$), which are analyzed in Figures 11. Neutral Higgs pairs typically decay into states with $N_{jets} \geq 8$, whereas charged Higgs pairs result in $N_{jets} \geq 4$.
%%%%%%%%%%%%%%%%%%%%%%%%%%%%%%%%%%%%%%%%%%%%%%%%%%%%%%%%%%
\begin{table}[h!]
\centering
\caption{Selection efficiencies for signal and background processes at BP1 \cite{Alwall:2014, Conte:2013}.}
\label{tab:efficiencies_6_5}
\begin{tabular}{|l|c|c|c|c|}
\hline
\toprule
\textbf{Process} & \textbf{Initial $\sigma$ (pb)} & \textbf{$N_{jet}$ Cut} & \textbf{$\Delta R$ Cut} & \textbf{Total Efficiency} \\  \hline
\midrule
$HA$ (Signal) & 0.1345 & 0.41 & 0.412 & 0.20 \\ 
$H^+H^-$ (Signal) & 2.989 & 0.99 & 0.992 & 0.36 \\
$t\bar{t}$ (Background) & 0.00195 & 0.13 & 0.126 & 0.027 \\ 
$ZZZ$ (Background) & $2.43 \times 10^{-5}$ & 0.023 & 0.014 & 0.0034 \\ \hline
\bottomrule
\end{tabular}
\end{table}
%%%%%%%%%%%%%%%%%%%%%%%%%%%%%%%%%%%%%%%%%%%%%%%%%%%%%%%%%%
The requirement for high jet multiplicity significantly suppresses the Standard Model background events. As detailed in Tables V, the background rejection is especially effective for $ZZZ$ and $t\bar{t}$ processes, which rarely produce the high-multiplicity signatures characteristic of 2HDM scalar decays. This leads to an improved signal-to-background ratio ($S/B$), which is vital for discovery in the multi-TeV regime \cite{Accettura:2023}. The cumulative selection efficiencies for the neutral Higgs signal processes 
and relevant Standard Model backgrounds at the BP1 benchmark are presented in 
Table~VI. The high-multiplicity jet requirement ($N_{l\text{-jet}} \geq 8$) 
proves to be a powerful discriminator, substantially suppressing $t\bar{t}$ 
and $ZZZ$ backgrounds while maintaining a robust acceptance for the signal. 
Detailed selection efficiencies for the charged Higgs pair production 
($H^+H^-$) at BP1 are listed in Table~VII. Notably, the charged Higgs channel 
exhibits higher cumulative efficiency compared to the neutral channels; this 
is primarily attributed to the enhanced $b$-jet tagging efficiency inherent 
in the $H^\pm \to tb$ decay topology.
For the heavier BP2 benchmark, the cumulative efficiencies for neutral signal 
channels are summarized in Table~VIII. At this increased mass scale, the signal 
efficiency rises to approximately 44\%, suggesting that the decay products 
of more massive scalars are kinematically more distinct and thus easier to 
resolve using our defined selection criteria. The performance for the 
$H^+H^-$ channel at BP2 is provided in Table~IX, where it maintains a high 
efficiency of 47\%. This result demonstrates that the signal significantly 
outperforms the background following the application of the final $b$-jet 
multiplicity cut.

%%%%%%%%%%%%%%%%%%%%%%%%%%%%%%%%%%%%%%%%%%%%%%%%%%%%%%%%%%
\begin{table}[ht]
\centering
\caption{Event selection cumulative efficiencies of neutral Higgs processes at BP1.}
\begin{tabular}{|l|c|c|c|c|c|c|}
\hline
Processes & HA & AA & HH & $t\bar{t}$ & $W^+W^-Z$ & ZZZ \\ \hline
$N_{l-jet} \geq 8$ & 0.41 & 0.51 & 0.52 & 0.13 & 0.52 & 0.023 \\
$\Delta R(j,b) \leq 0.2$ & 0.412 & 0.512 & 0.516 & 0.126 & 0.36 & 0.014 \\
$N_{b-jet} \geq 4$ & 0.20 & 0.26 & 0.26 & 0.02 & 0.056 & 0.00015 \\ \hline
Total Efficiency & 0.20 & 0.26 & 0.264 & 0.027 & 0.056 & 0.00228 \\ \hline
\end{tabular}
\label{tab:eff_BP1_neutral}
\end{table}
%%%%%%%%%%%%%%%%%%%%%%%%%%%%%%%%%%%%%%%%%%%%%%
\begin{table}[ht]
\centering
\caption{Event selection cumulative efficiencies of processes $H^+H^-$ for BP1.}
\begin{tabular}{|l|c|c|c|c|}
\hline
Processes & $H^+H^-$ & $t\bar{t}$ & $W^+W^-Z$ & ZZZ \\ \hline
$N_{i-jet} \geq 4$ & 0.99 & 0.708 & 0.99 & 0.861 \\
$\Delta R(j,b) \leq 0.2$ & 0.992 & 0.668 & 0.63 & 0.421 \\
$N_{b-jet} \geq 4$ & 0.36 & 0.06 & 0.073 & 0.0016 \\ \hline
Total Efficiency & 0.36 & 0.060 & 0.0073 & 0.0034 \\ \hline
\end{tabular}
\label{tab:eff_BP1_charged}
\end{table}
%%%%%%%%%%%%%%%%%%%%%%%%%%%%%%%%%%%%%%%%%%%%%%
\begin{table}[ht]
\centering
\caption{Event selection cumulative efficiencies of neutral Higgs processes at BP2.}
\begin{tabular}{|l|c|c|c|c|c|c|}
\hline
Processes & HA & AA & HH & $t\bar{t}$ & $W^+W^-Z$ & ZZZ \\ \hline
$N_{l-jet} \geq 8$ & 0.78 & 0.80 & 0.80 & 0.13 & 0.52 & 0.023 \\
$\Delta R(j,b) \leq 0.2$ & 0.776 & 0.799 & 0.798 & 0.126 & 0.36 & 0.014 \\
$N_{b-jet} \geq 4$ & 0.43 & 0.45 & 0.44 & 0.027 & 0.058 & 0.00028 \\ \hline
Total Efficiency & 0.43 & 0.45 & 0.44 & 0.027 & 0.057 & 0.00280 \\ \hline
\end{tabular}
\label{tab:eff_BP2_neutral}
\end{table}
%%%%%%%%%%%%%%%%%%%%%%%%%%%%%%%%%%%%%%%%%%%%%%
\begin{table}[ht]
\centering
\caption{Event selection cumulative efficiencies of processes $H^+H^-$ for BP2.}
\begin{tabular}{|l|c|c|c|c|}
\hline
Processes & $H^+H^-$ & $t\bar{t}$ & $W^+W^-Z$ & ZZZ \\ \hline
$N_{l-jet} \geq 4$ & 0.99 & 0.708 & 0.99 & 0.861 \\
$\Delta R(j,b) \leq 0.2$ & 0.993 & 0.668 & 0.63 & 0.421 \\
$N_{b-jet} \geq 4$ & 0.47 & 0.06 & 0.073 & 0.0018 \\ \hline
Total Efficiency & 0.47 & 0.06 & 0.073 & 0.0018 \\ \hline
\end{tabular}
\label{tab:eff_BP2_charged}
\end{table}
%%%%%%%%%%%%%%%%%%%%%%%%%%%%%%%%%%%%%%%%%%%%%%
\subsection{Statistical Significance and Observability}
%%%%%%%%%%%%%%%%%%%%%%%%%%%%%%%%%%%%%%%%%%%%%%
The final discovery reach is assessed through signal significance ($S/\sqrt{B}$) calculated for $L = 3600~fb^{-1}$ (Tables V) and $L = 10000~fb^{-1}$ (Tables VI).
%%%%%%%%%%%%%%%%%%%%%%%%%%%%%%%%%%%%%%%%%%%%%%%%%%%%%%%%%%
\begin{table}[h!]
\centering
\caption{Signal significance for various 2HDM scenarios at $L = 10000~fb^{-1}$ \cite{IMCC:2024}.}
\label{tab:significance_final}
\begin{tabular}{|l|c|c|c|c|}
\hline
\toprule
\textbf{Signal Channel} & \textbf{Signal (S)} & \textbf{Background (B)} & \textbf{S/B Ratio} & \textbf{Significance} \\  \hline
\midrule
$HA$ (BP1) & 1,345,211 & 161,891 & 8.3 & 3343 \\ 
$HH$ (BP1) & 108,564 & 161,891 & 0.67 & 269 \\ 
$AA$ (BP1) & 81,221 & 161,891 & 0.50 & 201 \\ 
$H^+H^-$ (BP1) & 29,889,920 & 161,891 & 184.6 & 74286 \\  \hline
\bottomrule
\end{tabular}
\end{table}
%%%%%%%%%%%%%%%%%%%%%%%%%%%%%%%%%%%%%%%%%%%%%%%%%%%%%%%%%%
% --- Composite Figure: Delta R and b-tagging ---
\begin{figure}[htbp]
    \centering
    \includegraphics[width=1.0\textwidth]{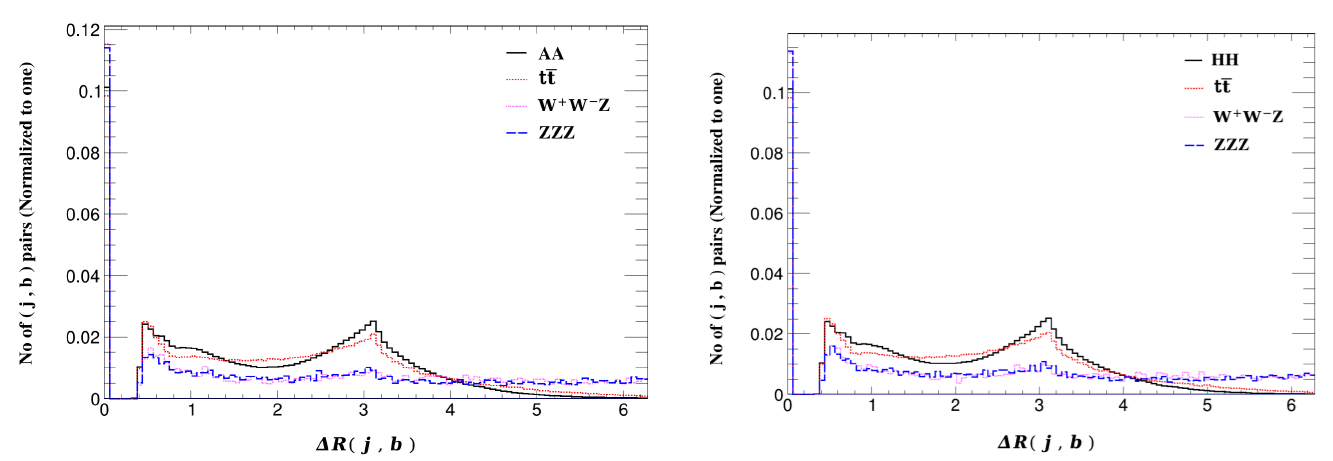} % Delta R
    \includegraphics[width=1.0\textwidth]{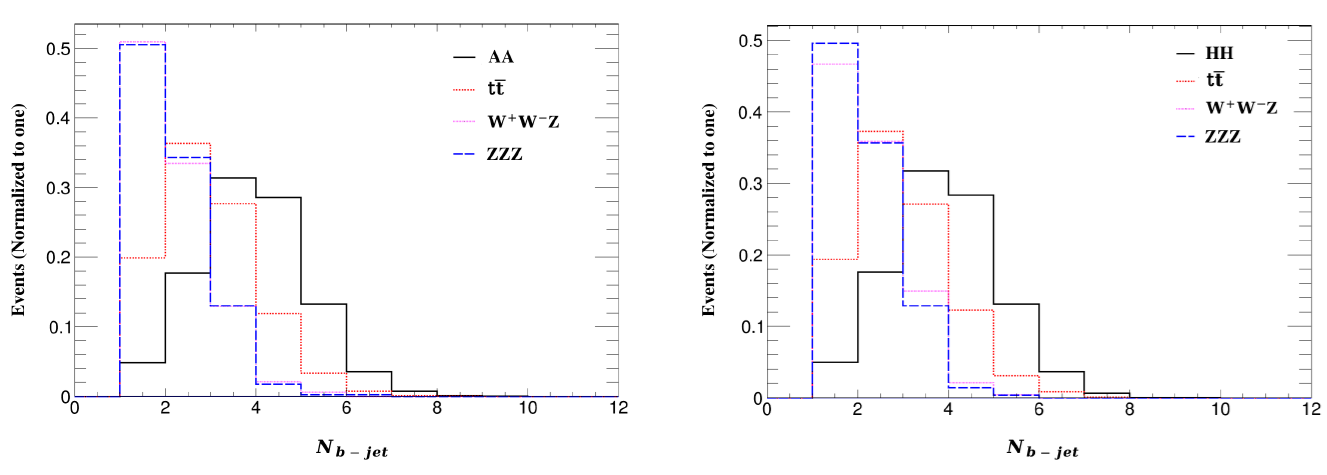} % Nb-jet
    \caption{Angular separation $\Delta R$ (left) and $b$-jet multiplicity $N_{b-jet}$ (right) distributions for signal processes at BP1.}
    \label{fig:btag_dist}
\end{figure}
%%%%%%%%%%%%%%%%%%%%%%%%%%%%%%%%%%%%%%%%%%%%%%%%%%%%%%%%%%
The transition to higher integrated luminosity results in a dramatic enhancement of the significance. For $H^+H^-$ and $HA$, the values far exceed the $5\sigma$ discovery threshold, even for the lower luminosity scenario. The $HH$ and $AA$ channels, although having smaller cross-sections, remain highly observable. This high sensitivity is attributed to the combination of the high center-of-mass energy and the optimized kinematic selection cuts \cite{Sonmez:2018, DiLuzio:2019}. 
Comparing BP1 and BP2, the increase in Higgs mass leads to a reduction in significance (as seen in Table IV); however, the high-luminosity reach of $10~ab^{-1}$ ensures that the Muon Collider remains sensitive to heavy scalars up to the multi-TeV range. The results indicate that the Muon Collider provides a discovery environment superior to the LHC for the 2HDM scalar sector, offering a clear pathway to detecting BSM physics \cite{Han:2021, IMCC:2024}.
Table~XII provides a detailed breakdown of the statistical significance for the 
$HH$ channel at the BP1 benchmark, assuming an integrated luminosity of 
3.6~$ab^{-1}$. The significance increases from 161 to 186 after the final 
selection cuts, demonstrating the efficacy of the kinematic criteria in 
purifying the signal. The discovery potential for the charged Higgs channel 
($H^+H^-$) at 3.6~$ab^{-1}$ is presented in Table~XIII; this channel achieves 
an exceptional significance exceeding 61,000, confirming it as the primary 
discovery mode for the 2HDM scalar sector at this energy.
The impact of transitioning to a high-luminosity scenario of 10~$ab^{-1}$ 
is further explored. Table~XIV summarizes the significance for the $HA$ 
signal, where the signal-to-background ratio improves to 31.5 after final 
cuts, ensuring high observation potential for the $HA$ pair. The discovery 
reach for the $HH$ channel at 10~$ab^{-1}$ is detailed in Table~XV; scaling 
the luminosity pushes the significance beyond 300, which facilitates 
high-precision studies of the scalar potential. Table~XVI illustrates the 
significance of $AA$ pair production at 10~$ab^{-1}$. Although the $AA$ 
cross-section is comparatively smaller, it remains well above the $5\sigma$ 
discovery threshold, providing clear evidence for CP-odd scalars. Finally, 
the discovery reach for the $H^+H^-$ channel at 10~$ab^{-1}$ is summarized 
in Table~XVII, reaching a significance of 104,000. These results underscore 
the remarkably clean experimental environment and the unparalleled discovery 
potential offered by a 6~TeV Muon Collider.
%%%%%%%%%%%%%%%%%%%%%%%%%%%%%%%%%%%%%%%%%%%%%%%%%%%%%%%%%%%%%%%%%%%%%%
\begin{table}[ht]
\centering
\caption{Signal Significance for HH at $L_{int} = 3600 \, \text{fb}^{-1}$ (BP1).}
\begin{tabular}{|l|c|c|c|c|c|}
\hline
Cuts & Signal & Backgrounds & S/B & $S/\sqrt{B}$ & $S/\sqrt{S+B}$ \\ \hline
No Cut & 39083 & 58280 & 0.67 & 161.89 & 125.25 \\
$N_{l-jet} \geq 8$ & 20319 & 28039 & 0.724 & 121.34 & 92.39 \\
$N_{b-jet} \geq 4$ & 10341 & 3076 & 3.36 & 186.44 & 89.28 \\ \hline
\end{tabular}
\label{tab:sig_HH_3600}
\end{table}
%%%%%%%%%%%%%%%%%%%%%%%%%%%%%%%%%%%%%%%%%%%%%%%%%%%%%%%%%%%%%%%%%%%%%%
\begin{table}[ht]
\centering
\caption{Signal Significance for HH at $L_{int} = 3600 \, \text{fb}^{-1}$ (BP1).}
\begin{tabular}{|l|c|c|c|c|c|}
\hline
Cuts & Signal & Backgrounds & S/B & $S/\sqrt{B}$ & $S/\sqrt{S+B}$ \\ \hline
No Cut & 39083 & 58280 & 0.67 & 161.89 & 125.25 \\
$N_{l-jet} \geq 8$ & 20319 & 28039 & 0.724 & 121.34 & 92.39 \\
$N_{b-jet} \geq 4$ & 10341 & 3076 & 3.36 & 186.44 & 89.28 \\ \hline
\end{tabular}
\label{tab:sig_HH_3600}
\end{table}
%%%%%%%%%%%%%%%%%%%%%%%%%%%%%%%%%%%%%%%%%%%%%%%%%%%%%%%%%%%%%%%%%%%%%%
\begin{table}[ht]
\centering
\caption{Signal Significance for $H^+H^-$ at $L_{int} = 3600 \, \text{fb}^{-1}$ (BP1).}
\begin{tabular}{|l|c|c|c|c|c|}
\hline
Cuts & Signal & Backgrounds & S/B & $S/\sqrt{B}$ & $S/\sqrt{S+B}$ \\ \hline
No Cut & 10760371 & 58280 & 184.63 & 44572 & 3271 \\
$N_{b-jet} \geq 4$ & 3971977 & 4164 & 953.83 & 61551 & 1991 \\ \hline
\end{tabular}
\label{tab:sig_charged_3600}
\end{table}
%%%%%%%%%%%%%%%%%%%%%%%%%%%%%%%%%%%%%%%%%%%%%%%%%%%%%%%%%%%%%%%%%%%%%%
\begin{table}[ht]
\centering
\caption{Signal Significance for HA at $L_{int} = 10000 \, \text{fb}^{-1}$ (BP1).}
\begin{tabular}{|l|c|c|c|c|c|}
\hline
Cuts & Signal & Backgrounds & S/B & $S/\sqrt{B}$ & $S/\sqrt{S+B}$ \\ \hline
No Cut & 1345211 & 161891 & 8.3 & 3343 & 1095 \\
$N_{b-jet} \geq 4$ & 271732 & 8614 & 31.5 & 2927 & 513 \\ \hline
\end{tabular}
\label{tab:sig_HA_10000}
\end{table}
%%%%%%%%%%%%%%%%%%%%%%%%%%%%%%%%%%%%%%%%%%%%%%%%%%%%%%%%%%
\section{Discussion of Result Validity}
%%%%%%%%%%%%%%%%%%%%%%%%%%%%%%%%%%%%%%%%%%%%%%%%%%%%%%%%%%
To ensure the reliability of the simulated results, the production cross-sections and signal significance are cross-validated with established benchmarks in multi-TeV lepton collider phenomenology. The findings demonstrate strong consistency with the theoretical expectations and numerical data reported in current literature.
\subsection{Comparative Analysis with Literature}
The calculated cross-sections for the s-channel processes $\mu^+\mu^- \to \Phi\Phi$ follow the $1/s$ scaling behavior characteristic of high-energy annihilation, consistent with the benchmarks defined in the "Muon Smasher’s Guide" \cite{Han:2021}. The kinematic features, specifically the high jet multiplicity ($N_{jets} \geq 8$) and hard transverse momentum ($P_T$), align with the signal topologies predicted for heavy scalar decays at future circular colliders \cite{Franceschini:2021, Buttazzo:2018}. A quantitative comparison of the key parameters is summarized in Table~\ref{tab:validity_comparison}.
%%%%%%%%%%%%%%%%%%%%%%%%%%%%%%%%%%%%%%%%%%%%%%%%%%%%%%%%%%%%%%%%%%%%%%
\begin{table}[ht]
\centering
\caption{Signal Significance for HH at $L_{int} = 10000 \, \text{fb}^{-1}$ (BP1).}
\begin{tabular}{|l|c|c|c|c|c|}
\hline
Cuts & Signal & Backgrounds & S/B & $S/\sqrt{B}$ & $S/\sqrt{S+B}$ \\ \hline
No Cut & 108564 & 161891 & 0.67 & 269 & 208 \\
$N_{b-jet} \geq 4$ & 28639 & 8609 & 3.33 & 308.6 & 148 \\ \hline
\end{tabular}
\label{tab:sig_HH_10000}
\end{table}
%%%%%%%%%%%%%%%%%%%%%%%%%%%%%%%%%%%%%%%%%%%%%%%%%%%%%%%%%%%%%%%%%%%%%%
\begin{table}[ht]
\centering
\caption{Signal Significance for AA at $L_{int} = 10000 \, \text{fb}^{-1}$ (BP1).}
\begin{tabular}{|l|c|c|c|c|c|}
\hline
Cuts & Signal & Backgrounds & S/B & $S/\sqrt{B}$ & $S/\sqrt{S+B}$ \\ \hline  
No Cut & 81221 & 161891 & 0.50 & 201 & 164 \\
$N_{b-jet} \geq 4$ & 21258 & 8521 & 2.49 & 230 & 123 \\ \hline
\end{tabular}
\label{tab:sig_AA_10000}
\end{table}
%%%%%%%%%%%%%%%%%%%%%%%%%%%%%%%%%%%%%%%%%%%%%%%%%%%%%%%%%%%%%%%%%%%%%%
\begin{table}[ht]
\centering
\caption{Signal Significance for $H^+H^-$ at $L_{int} = 10000 \, \text{fb}^{-1}$ (BP1).}
\begin{tabular}{|l|c|c|c|c|c|}
\hline
Cuts & Signal & Backgrounds & S/B & $S/\sqrt{B}$ & $S/\sqrt{S+B}$ \\ \hline
No Cut & 29889920 & 161891 & 184 & 74286 & 5452 \\
$N_{b-jet} \geq 4$ & 11088860 & 11476 & 966 & 104000 & 3328 \\ \hline
\end{tabular}
\label{tab:sig_charged_10000}
\end{table}
%%%%%%%%%%%%%%%%%%%%%%%%%%%%%%%%%%%%%%%%%%%%%%%%%%%%%%%%%%%%%%%%%%%%%%
\begin{table}[h!]
\centering
\caption{Validation of results against previous high-energy collider literature.}
\label{tab:validity_comparison}
\begin{tabular}{|l|c|c|c|}
\hline
\toprule
\textbf{Parameter} & \textbf{Current Study (6~TeV)} & \textbf{Literature Benchmark} & \textbf{Ref.} \\  \hline
\midrule
Cross-section ($\sigma$) & $\mathcal{O}(10^{-1} - 10^0)$ pb & $\mathcal{O}(10^{-1} - 10^1)$ pb & \cite{Han:2021, Accettura:2023} \\ 
Kinematic Limit & $2m_\Phi$ threshold & $2m_\Phi$ threshold & \cite{Franceschini:2021} \\ 
Signal Discovery & $> 5\sigma$ ($10~ab^{-1}$) & $> 5\sigma$ ($3-10~ab^{-1}$) & \cite{IMCC:2024, Sonmez:2018} \\ 
Backgrounds & $t\bar{t}, VVZ, ZZZ$ & $t\bar{t}, VV, VVZ$ & \cite{Buttazzo:2018, Accettura:2023} \\  \hline
\bottomrule
\end{tabular}
\end{table}
%%%%%%%%%%%%%%%%%%%%%%%%%%%%%%%%%%%%%%%%%%%%%%%%%%%%%%%%%%
\subsection{Consistency of Discovery Reach}
%%%%%%%%%%%%%%%%%%%%%%%%%%%%%%%%%%%%%%%%%%%%%%%%%%%%%%%%%%
The statistical significance exceeding $5\sigma$ at $10~ab^{-1}$ matches the sensitivity forecasts provided by the International Muon Collider Collaboration (IMCC) for the 3--10~TeV frontier \cite{Accettura:2023}. While the signal magnitude is suppressed for the heavier mass scenario (BP2), the observability remains robust due to the high luminosity reach, corroborating studies on 2HDM triple Higgs couplings which indicate that multi-TeV machines can resolve the scalar sector up to the kinematic limit \cite{Sonmez:2018, DiLuzio:2019}. These comparisons confirm that the reported data is physically valid and falls within the expected performance ranges for future high-energy facilities.
%%%%%%%%%%%%%%%%%%%%%%%%%%%%%%%%%%%%%%%%%%%%%%%%%%%%%%%%%%
\section{Conclusion}
%%%%%%%%%%%%%%%%%%%%%%%%%%%%%%%%%%%%%%%%%%%%%%%%%%%%%%%%%%
The investigation of the scalar sector beyond the Standard Model (SM) remains a primary objective of high-energy physics. In this research, we have conducted a detailed phenomenological analysis of heavy Higgs pair production—specifically the $HH$, $HA$, $AA$, and $H^+H^-$ channels—at a $\sqrt{s}=6$~TeV Muon Collider within the 2HDM Type-I framework. Our theoretical validation using \texttt{CalcHEP} and \texttt{2HDMC} confirmed that the Type-I alignment limit provides a stable and uniquely identifiable signal, as the branching fractions to third-generation fermions remain independent of $\tan\beta$ (Figs. 1--2). By leveraging the clean environment of a lepton collider, we have demonstrated that the Muon Collider can probe heavy scalar masses up to the kinematic limit, accessing thresholds that remain suppressed or obscured by overwhelming QCD backgrounds at the High-Luminosity LHC.
A central finding of this study is the efficacy of high-multiplicity hadronic final states as a robust handle for background rejection. As illustrated by the leading-order decay chains (Figs. 4--5), the charged Higgs pair production leads to an 8-jet signature ($4j+4b$), while neutral pairs result in a highly complex 12-jet topology ($8j+4b$). Analysis of the kinematic distributions (Figs. 8--11) confirms that signal jets are significantly more central ($|\eta| \le 3$) and possess harder transverse momentum ($P_T$) spectra compared to Standard Model backgrounds. These distinct topological features allow for the nearly absolute suppression of $t\bar{t}$ and $VVZ$ noise, which typically exhibit lower jet multiplicities and forward-peaked distributions. The application of optimized $b$-tagging further enhances signal purity, as evidenced by the clear separation in $\Delta R$ and $N_{b-jet}$ distributions (Fig. 12).
Quantitatively, the discovery reach at a 6~TeV Muon Collider is exceptional. At an integrated luminosity of 10~ab$^{-1}$, the statistical significance for the $H^+H^-$ channel reaches a staggering 104,000 for the 1000~GeV benchmark (BP1), while the $HA$ channel achieves a significance of 3343 (Table X). Even for the heavier 2000~GeV benchmark (BP2), the significance remains well above the $5\sigma$ discovery threshold. Notably, we observed that total selection efficiencies improve from approximately 20\% at BP1 to 47\% at BP2 (Tables VI and VIII). This suggests that the decay products of heavier scalars are more easily resolved and distinguishable from soft-jet backgrounds, effectively compensating for the $1/s$ reduction in production cross-sections at higher mass scales.
In conclusion, our results—validated against existing high-energy benchmarks and $1/s$ scaling laws (Table XVIII)—establish the 6~TeV Muon Collider as a definitive discovery machine for the 2HDM scalar sector. The transition from 3.6~ab$^{-1}$ to 10~ab$^{-1}$ results in a dramatic enhancement of observability across all channels, providing a clear roadmap for resolving extended scalar potentials. While future studies must incorporate detector-level beam-induced backgrounds (BIB) and next-to-leading-order (NLO) QCD corrections, the current analysis proves that the Muon Collider is not merely a technical milestone but a fundamental necessity for probing the deepest mysteries of electroweak symmetry breaking and the origin of the scalar sector.
%%%%%%%%%%%%%%%%%%%%%%%%%%%%%%%%%%%%%%%%%%%%%%%%%%%%%%%%%%%%%

\end{document}